\documentstyle[aps,epsf]{revtex}  
\begin{document} 
\def\Paris{Par\'\i{}s}
\def\Perez{P\'erez}
\def\Astrofisica{Astrof\'\i{}sica}
\def\Astrobiologia{Astrobiolog\'{\i}a}
\def\Fisica{F\'\i{}sica}
\def\Torrejon{Torrej\'on}
\title{ 
Effective potential for the massless KPZ equation
} 
\author{ 
David Hochberg$^{+,*}$, Carmen Molina--\Paris$^{++}$,  
Juan \Perez--Mercader$^{+++,*}$, and Matt Visser$^{++++}$ 
} 
\address{ 
$^{+,+++}$Laboratorio de \Astrofisica\ Espacial y \Fisica\ 
Fundamental, Apartado 50727, 28080 Madrid, Spain\\ 
$^{++}$Theoretical Division, Los Alamos National Laboratory, 
Los Alamos, New Mexico 87545, USA\\ 
$^{++++}$Physics Department, Washington University, 
Saint Louis, Missouri 63130-4899, USA\\  
$^{*}$Centro de \Astrobiologia\ 
(Associate Member NASA Astrobiology Institute),
INTA, Ctra. Ajalvir, Km. 4, 28850 \Torrejon, Madrid, Spain}
\date{26 April 1999; \LaTeX-ed \today} 
\maketitle 
\bigskip

{\small 

{\bf Abstract:} In previous work we have developed a general method
for casting a classical field theory subject to Gaussian noise (that
is, a stochastic partial differential equation---SPDE) into a
functional integral formalism that exhibits many of the properties
more commonly associated with quantum field theories (QFTs). In
particular, we demonstrated how to derive the one-loop effective
potential. In this paper we apply the formalism to a specific field
theory of considerable interest, the massless KPZ equation (massless
noisy vorticity-free Burgers equation), and analyze its behaviour in
the ultraviolet (short-distance) regime.  When this field theory is
subject to white noise we can calculate the one-loop effective
potential and show that it is one-loop ultraviolet renormalizable in
1, 2, and 3 space dimensions, and fails to be ultraviolet
renormalizable in higher dimensions.  We show that the one-loop
effective potential for the massless KPZ equation is closely related
to that for $\lambda \phi^4$ QFT. In particular we prove that the
massless KPZ equation exhibits one-loop dynamical symmetry breaking
(via an analog of the Coleman--Weinberg mechanism) in 1 and 2 space
dimensions, and that this behaviour does not persist in 3 space
dimensions.

\bigskip


} 

\newcommand{\Str}{\mathop{\mathrm{Str}}} 
\newcommand{\tr}{\mathop{\mathrm{tr}}} 
\newcommand{\define}{\mathop{\stackrel{\rm def}{=}}}
\newcommand{\Tr}{\mathop{\mathrm{Tr}}}
\def\d{{\mathrm d }}
\def\implies{\Rightarrow}
\def\half{ {\scriptstyle{1\over2}} }
\def\quarter{ {\scriptstyle{1\over4}} }
\def\A{ {\cal A} }
\def\sign{ \hbox{sign} }
\def\Quiros{Quir\'os}
\def\Leon{Le\'on}
\section{Introduction}

In a companion paper~\cite{HMPV-spde}, we discussed classical field
theories subject to stochastic noise $\eta(\vec x, t)$, described by
the equation
\begin{equation}
D \phi(\vec x,t) = F[\phi(\vec x,t)] + \eta(\vec x,t).
\end{equation}
Here $D$ is any linear differential operator, involving arbitrary time
and space derivatives, that does {\em not} explicitly involve the
field $\phi$.  The function $F[\phi]$ is any forcing term, generally
nonlinear in the field $\phi$. These stochastic partial differential
equations (SPDEs) can be studied using a functional integral formalism
which makes manifest the deep connections with quantum field theories
(QFTs). Provided the noise is translation-invariant and Gaussian, it
is possible to split its two-point function into an {\em amplitude}
${\cal A}$ and a {\em shape} function $g_2(x,y)$, as follows
\begin{equation}
G_\eta(x,y) \define \A \; g_2(x-y),
\end{equation}
with the {\em convention} that
\begin{equation}
\int  \d^d \vec x\; \d t \; g_2^{-1}(\vec x,t)
\; = \; 1
\; = \; \tilde g_2^{-1}(\vec k=\vec 0,\omega=0).
\end{equation}
We showed that the one-loop effective potential for homogeneous and
static fields is~\cite{HMPV-spde}
\begin{eqnarray}
\label{E:general}
{\cal V}[\phi;\phi_0] &=& 
\half F^2[\phi]
+ \half \A \int {\d^d \vec k \; \d \omega\over (2\pi)^{d+1}}
\ln  
\left[ 1 + {\tilde g_2{}(\vec k,\omega)
F[\phi] {\delta^2 F\over\delta\phi\;\delta\phi} 
\over
\left( D^\dagger(\vec k,\omega)  - {\delta F\over \delta\phi}^\dagger \right)
\left( D(\vec k,\omega) - {\delta F\over \delta\phi} \right)}
\right]
- \left( \phi \to \phi_0 \right)
+ O(\A^2).
\end{eqnarray}
Here $\phi_0$ is any convenient background field. In the absence of
symmetry breaking it is most convenient to pick $\phi_0=0$, but we
reserve the right to make other choices when appropriate. Equation
(\ref{E:general}) is qualitatively similar to the one-loop effective
potential for a self-interacting scalar
QFT~\cite{Weinberg,Zinn-Justin}:
\begin{eqnarray}
{\cal V}[\phi;\phi_0] &=& 
V(\phi)
+ \half \hbar \int {\d^d \vec k \; \d \omega\over (2\pi)^{d+1}}
\ln  
\left[ 1 + { {\delta^2 V\over\delta\phi\;\delta\phi} 
\over
\omega^2 + \vec k^2 + m^2
}
\right]
- \left( \phi \to \phi_0 \right)
+ O(\hbar^2).
\end{eqnarray}
The effective potential is not only a formal mathematical tool, but it
also has a deep physical meaning. In fact, in previous
work~\cite{HMPV-spde} we demonstrate that the effective potential for
SPDEs inherits many of the physical features and information content
of the effective potential for QFTs, and that searching for minima of
the SPDE effective potential provides information about ``ground
states'' of such SPDEs. These ground states play an important role in
the study of symmetry breaking and the onset of pattern formation and
structure.  In this paper we apply this formalism to the specific case
of the massless KPZ equation (equivalent to the massless noisy
vorticity-free Burgers
equation)~\cite{Frisch,KPZ,MHKZ,Sun-Plischke,Frey-Tauber}
\begin{equation}
\left({\partial\over\partial t} - \nu \vec\nabla^2\right) \phi = 
F_0+{\lambda\over2} (\vec \nabla\phi)^2 + \eta,
\end{equation}
and use the formalism to investigate the ultraviolet (short-distance)
properties of the system.

In the fluid dynamical interpretation ({\em i.e.}, the vorticity-free
Burgers equation) the fluid velocity is taken to be $\vec v = -
\vec\nabla \phi$.  In this representation the KPZ equation is used as
a model for turbulence~\cite{Frisch}, structure development in the
early universe~\cite{Structure}, driven diffusion, and flame
fronts~\cite{MHKZ}.  In the surface growth interpretation, $\phi(\vec
x,t)$ is taken to be the height of the surface (typically defined over
a two-dimensional plane)~\cite{KPZ}.  In this interpretation the
massless KPZ equation is a natural nonlinear extension of the
Edwards--Wilkinson (EW) model~\cite{Barabasi}.

An explicit tadpole term ($F_0$) is included as we will soon see that
it is a necessary ingredient in completing the renormalization
program. (In QFTs a dynamically generated field-independent constant
term in the equations of motion is often called a tadpole, though if
such a term arises from the tree-level physics it is called an
external current. In this KPZ context we prefer to use the word
``tadpole'' since $F_0$ will have to be renormalized, which makes the
nomenclature ``external current'' inappropriate.)  After
renormalization, we will fix the value of the {\em renormalized} value
of $F_0$ by using the symmetries of the KPZ equation.

Some additional comments regarding the {\em raison-d'etre} of the
tadpole may prove helpful.  It is well known that both the
Edwards--Wilkinson and KPZ equations may be written with or without
such a constant (the tadpole) and that a finite constant term simply
represents a change in the average velocity of the surface with
respect to the laboratory frame of reference, and so may be chosen at
will. On the other hand, as we will see below, it is crucial to
include a bare tadpole in the KPZ equation in order to be able to
consistently carry out the ultraviolet renormalization (at
one-loop). Nevertheless, the renormalized tadpole may be chosen at
will, and can be set to any convenient value after
renormalization. (The convenient value we will finally adopt will be
based on symmetry arguments and will be chosen to eliminate any
spurious motion of the background field.)  The need for a ``bare''
tadpole is not just an artifact of the effective potential
renormalization but (as we will demonstrate elsewhere) is also
required to carry out the one-loop ultraviolet renormalization of the
full effective action.

There are {\em two} important symmetries of the KPZ equation that are
relevant for our analysis. The first is the symmetry under the
transformation
\begin{eqnarray}
\phi &\to& \phi + c(t), \\
F_0  &\to& F_0 + {d c(t)\over dt}. 
\end{eqnarray}
In the fluid dynamics interpretation this symmetry amounts to a
``gauge transformation'' of the scalar field $\phi$ that does not
change the physical velocity ($\vec v = - \vec \nabla \phi$). In the
surface growth interpretation this symmetry corresponds to choosing a
different coordinate system that moves vertically at a speed $dc/dt$
with respect to the initial coordinate system, and so can be thought
of as a type of Galilean invariance (type I) for the KPZ equation. This
transformation is a symmetry of the KPZ equation for arbitrary noise.

The second symmetry we consider holds under more restrictive conditions. 
Consider the transformation
\begin{eqnarray}
\vec x &\to& \vec x' = \vec x - \lambda \; \vec \epsilon \; t, \\
t &\to& t' = t, \\
\phi(\vec x,t) &\to& \phi'(\vec x',t') 
= \phi(\vec x,t) - \vec \epsilon \cdot \vec x.
\end{eqnarray}
In the fluid dynamics interpretation this symmetry is equivalent to a
Galilean transformation of the fluid velocities
\begin{equation}
\vec v \to \vec v' = \vec v - \vec \epsilon,
\end{equation}
and so can also be thought of as a type of Galilean invariance (type
II).  In the surface growth interpretation this symmetry amounts to
choosing a different coordinate system that is tilted at an angle to
the vertical, with
\begin{equation}
\tan(\theta) = ||\,\vec \epsilon\, ||,
\end{equation}
and for this reason this transformation is often referred to as tilt
invariance. While this type II Galilean invariance is an exact
invariance of the zero-noise KPZ equation, it is important to keep in
mind that once noise is added to the system, this transformation will
remain a symmetry only if the noise is translation-invariant and {\em
temporally white}. This can be seen by first looking at the noise two
point function
\begin{equation}
G_\eta(\vec x_1,t_1; \vec x_2,t_2) = \A \; g_2(\vec x_1-\vec x_2, t_1-t_2),
\end{equation}
then considering
\begin{eqnarray}
\vec x_1 - \vec x_2 &\to& \vec x_1' - \vec x_2' 
= \vec x_1 - \vec x_2 - \lambda \; \vec \epsilon \; (t_1 - t_2),
\\
t_1 - t_2 &\to& t_1' - t_2' = t_1-t_2,
\end{eqnarray}
and finally noting that the noise two-point function is invariant if
and only if its support is limited by the constraint $t_1=t_2$, that
is,
\begin{equation}
G_\eta(\vec x_1,t_1; \vec x_2,t_2) = 
\A \; \hat g_2(\vec x_1-\vec x_2) \; \delta(t_1-t_2).
\end{equation}
But this is the {\em definition} of translation-invariant temporally
white noise.

We will use these two symmetries extensively in the body of the paper:
the type I Galilean invariance is used to guarantee that the
background field $\phi_0$ is kept stationary, even in the presence of
interactions and nonlinearities, and the type II Galilean invariance
is similarly used to guarantee that the average slope of the
background field is zero. It is best (in fact essential) to do this
only after the renormalization program has been completed, so for
the time being we will explicitly keep track of both the tadpole term
$F_0$ and the background field $\phi_0$.

Applying the formalism developed in~\cite{HMPV-spde} to the massless
KPZ equation, we demonstrate that the effective potential is one-loop
ultraviolet renormalizable in 1, 2, and 3 space dimensions (and is not
ultraviolet renormalizable in 4 or higher space dimensions).  We will
discover a formal relationship between the one-loop effective
potential for the massless KPZ equation in $d+1$ dimensions ($d$ space
and 1 time dimensions) and that of the massless $\lambda \phi^4$ QFT
in $d+2$ Euclidean dimensions, and show that the massless KPZ equation
exhibits many of the properties seen in massless $\lambda \phi^4$ QFT.
In particular, we will see that the stochastic system undergoes
dynamical symmetry breaking (DSB) in 1 and 2 space dimensions, and no
symmetry breaking in 3 dimensions.  This DSB is due to an analog of
the Coleman--Weinberg mechanism of QFT. In 2 space dimensions the
presence of a short-distance (ultraviolet) logarithmic divergence
implies the running of the coupling constant $\lambda$ with the energy
scale, and the existence of a non-zero beta function for this
coupling.  We feel that the unexpected presence of DSB in 1 and 2
space dimensions is a matter of deep importance, and is something that
would be very difficult to deduce by any other means.

\section{Effective Potential: massless KPZ}

The massless KPZ equation (massless noisy vorticity-free Burgers
equation) is~\cite{Frisch,KPZ,MHKZ,Sun-Plischke,Frey-Tauber}
\begin{equation}
\label{E:KPZ}
\left({\partial\over\partial t} - \nu \vec\nabla^2\right) \phi = 
F_0 +{\lambda\over2} (\vec\nabla\phi)^2  + \eta.
\end{equation}
We have introduced an explicit ``bare'' tadpole term $F_0$ as remarked
above.  If we now restrict attention to homogeneous and static fields,
$\phi(\vec x,t) \to \phi_{\mathrm homogeneous-static}$, and
$\phi_0(\vec x,t) \to (\phi_0)_{\mathrm homogeneous-static}$, then
from equation (\ref{E:general}) it is easy to see that ${\cal
V}[\phi,\phi_0] \equiv 0$, and so the one-loop effective potential is
uninteresting.  In order to see this, note that for a homogeneous
static field configuration
\begin{equation}
F[\phi] \to F_0; 
\qquad
{\delta F\over\delta\phi(x)} \to 0;
\qquad
{\delta^2 F \over \delta\phi(x) \delta\phi(y)}
= + \lambda \vec \nabla_x \cdot \vec \nabla_y \delta^d (\vec x, \vec y)
\to - \lambda \vec \nabla_x^2 
\to + \lambda \vec k^2.
\end{equation}
Thus the integrand appearing in (\ref{E:general}) is independent of
the field $\phi$, and ${\cal V}[\phi;\phi_0]$ is zero as asserted.
(This simplification does not hold for the massive noisy
vorticity-free Burgers equation [massive KPZ equation] where the
driving force is replaced by $F[\phi] \to F[\phi] - m^2 \phi$. We have
calculated ${\cal V}[\phi;\phi_0]$ explicitly for this system and
shown it is non-zero.  We do not report the details here, as this is a
simple exercise and the result is of limited physical interest.)

What {\em is} physically interesting on the other hand, is to study
the {\em massless} KPZ equation and to consider a linear and static
field configuration
\begin{equation}
\phi = - \vec v \cdot \vec x,
\end{equation}
where $\vec v$ is now a constant vector. Notice that for this choice
one has $D \phi = 0$. In the hydrodynamic interpretation of the
massless KPZ equation this corresponds to a constant velocity flow:
$\vec v = - \vec\nabla\phi$.  In the surface growth interpretation,
$\vert \vert \, \vec v \, \vert \vert$ corresponds to a constant slope
of the surface~\cite{KPZ}. There is an instructive analogy with QED
here: taking a constant vector potential in QED is relatively
uninteresting, it corresponds to zero electromagnetic field strength
and can be gauged away (modulo topological constraints).  On the other
hand, a constant electromagnetic field strength can be described by a
linear vector potential, and leads to such useful quantities as the
Euler--Heisenberg effective potential and the Schwinger effective
Lagrangian for QED~\cite{Weinberg,Zinn-Justin}.

Inspection of the derivation contained in reference~\cite{HMPV-spde}
reveals that although the effective potential (\ref{E:general}) was
originally defined for homogeneous and static fields, for the massless
KPZ equation (\ref{E:KPZ}), it continues to make sense for linear and
static fields. The fact that for the massless KPZ equation $F[\phi]$
is position independent for these linear static fields is essential to
this observation.  For such a field configuration, $\phi = - \vec v
\cdot \vec x$, we get
\begin{equation}
F[\phi] \to F_0 + {\lambda\over2} v^2 ; 
\qquad
{\delta F\over\delta\phi(x)} \to 
-\lambda \vec v \cdot \vec \nabla_x ;
\qquad
{\delta^2 F \over \delta\phi(x) \delta\phi(y)}
= + \lambda \vec \nabla_x \cdot \vec \nabla_y
\to - \lambda \vec \nabla_x^2 
\to + \lambda \vec k^2.
\end{equation}
The zero-loop effective potential is
\begin{equation}
{\cal V}_{\mathrm zero-loop}[v;v_0] = 
\half \left[
\left( F_0 + \half{\lambda} v^2 \right)^2 - 
\left( F_0 + \half{\lambda} v_0^2 \right)^2 
\right],
\end{equation}
where the background field is $\phi_0=-\vec v_0 \cdot \vec x$.  Note
that this zero-loop effective potential is formally equivalent to that
of $\lambda \phi^4$ QFT---with the velocity $v$ playing the role of
the quantum field $\phi_{\mathrm QFT}$.  Even at zero loops (tree
level) we see that if we were to have $F_0<0$ the effective potential
would take on the ``Mexican hat'' form, so that the onset of
spontaneous symmetry breaking (SSB) would not be at all
unexpected~\cite{Weinberg,Zinn-Justin}.  However, as previously
mentioned, the renormalized value of $F_0$ is not physically relevant
and can always be changed by a type I Galilean transformation. We will
soon see that SSB is not a feature of the KPZ equation.  Instead we
encounter a much more subtle effect: the onset of dynamical symmetry
breaking (DSB) which we have detected via a one-loop computation.

We start the one-loop calculation by noting that
\begin{equation}
D - {\delta F\over\delta \phi} = 
\partial_t + \lambda \; \vec v \cdot \vec \nabla - \nu \vec \nabla^2
\qquad \to \qquad -i\omega +i\lambda \vec v \cdot \vec k + \nu \vec k^2,
\end{equation}
while for the adjoint quantity
\begin{equation}
D^\dagger - {\delta F\over\delta \phi}^\dagger = 
-\partial_t - \lambda \; \vec v \cdot \vec \nabla - \nu \vec \nabla^2
\qquad \to \qquad +i\omega -i\lambda \vec v \cdot \vec k + \nu \vec k^2,
\end{equation}
so that
\begin{equation}
\left(D^\dagger -{\delta F\over \delta\phi}^\dagger\right)
\left(D - {\delta F \over\delta\phi}\right)
= 
-(\partial_t + \lambda \vec v \cdot \vec \nabla)^2 + \nu^2 (\vec \nabla^2)^2
\qquad \to \qquad
(\omega - \lambda \; \vec v \cdot \vec k)^2 + \nu^2 (\vec k^2)^2.
\end{equation}
Using this we specialize equation (\ref{E:general}) to 
\begin{eqnarray}
{\cal V}[v;v_0] &=& 
\half(F_0+\half{\lambda} v^2)^2
+\half \A \int {\d^d \vec k \; \d \omega\over (2\pi)^{d+1}}
\ln  
\left[ 1 + {\tilde g_2{}(\vec k,\omega) 
\lambda (F_0+\half{\lambda} v^2) \vec k^2
\over 
(\omega - \lambda \vec v \cdot \vec k)^2 + \nu^2 (\vec k^2)^2}
\right]
- \left( \vec v \to \vec v_0 \right)
+ O(\A^2).
\end{eqnarray}
Equivalently
\begin{eqnarray}
{\cal V}[v;v_0] &=& 
\half(F_0+\half{\lambda} v^2)^2
+ \half \A \int {\d^d \vec k \; \d \omega\over (2\pi)^{d+1}}
\ln  
\left[ {
(\omega - \lambda \vec v \cdot \vec k)^2 + \nu^2 (\vec k^2)^2 + 
\tilde g_2{}(\vec k,\omega) 
\lambda (F_0+\half{\lambda} v^2) \vec k^2
\over 
(\omega- \lambda \vec v \cdot \vec  k)^2 + \nu^2 (\vec k^2)^2}
\right]
\nonumber\\
&& 
- \left( \vec v \to \vec v_0 \right)
+ O(\A^2).
\end{eqnarray}
This is as far as we can go {\em without making further assumptions
about the noise}. For instance, one very popular choice is {\em
temporally white}, which means delta function correlated in time so
that $\tilde g_2(\vec k,\omega)\to \tilde g_2(\vec k)$ is a function
of $\vec k$ only. We can shift the integration variable $\omega$ to
$\omega - \lambda \vec v \cdot \vec k$.  (The frequency integral is
convergent.) The resulting integral becomes
\begin{eqnarray}
{\cal V}[v;v_0] &=& 
\half  (F_0+\half{\lambda} v^2)^2 
+ \half \A \int {\d^d \vec k \; \d \omega\over (2\pi)^{d+1}}
\ln  
\left[ {\omega^2 + \nu^2 (\vec k^2)^2 
+\tilde g_2{}(\vec k) 
\lambda (F_0+\half{\lambda} v^2) \vec k^2
\over 
\omega^2 + \nu^2 (\vec k^2)^2 }
\right]
\nonumber\\
&&
- \left( \vec v \to \vec v_0 \right) 
+ O(\A^2).
\end{eqnarray}
The fact that after this change of variables the denominator of the
logarithm in the two integrands is independent of the fields $\vec v$
and $\vec v_0$ is actually a surprisingly deep result, related to the
fact that the Jacobian functional determinant encountered
in~\cite{HMPV-spde} is a field-independent constant for the KPZ
equation.  We now make use of the integral
\begin{equation}
\int_{-\infty}^{+\infty} \d \omega 
\ln \left( {\omega^2 + X^2\over \omega^2 + Y^2} \right) = 
2\pi \left( X - Y \right),
\end{equation}
to deduce
\begin{eqnarray}
{\cal V}[v;v_0] &=& 
\half \left[
(F_0+\half{\lambda} v^2)^2 - (F_0+\half{\lambda} v_0^2)^2 
\right]
\nonumber\\
&&
+ \half \A \int {\d^d \vec k  \over (2\pi)^{d}} 
\left\{
\sqrt{   \nu^2 (\vec k^2)^2 
+ \tilde g_2{}(\vec k) 
\lambda (F_0+\half{\lambda} v^2) \vec k^2
} 
- \sqrt{ \nu^2 (\vec k^2)^2 
+ \tilde g_2{}(\vec k)
\lambda (F_0+\half{\lambda} v_0^2)  \vec k^2}
\right\}
\nonumber\\
&&
+ O(\A^2).
\end{eqnarray}
This may be simplified by extracting an explicit factor of $\nu^2 \vec
k^2$.  If we do so, the one-loop effective potential becomes
\begin{eqnarray}
\label{E:eff-potl}
{\cal V}[v;v_0] &=&
 \half \left[ (F_0+\half{\lambda} v^2)^2 -
(F_0+\half{\lambda} v_0^2)^2 \right] 
\nonumber\\ 
&& + \half \A \nu
\int {\d^d \vec k |k| \over (2\pi)^{d}}
 \left\{ \sqrt{ \vec k^2 +
\tilde g_2{}(\vec k) {\lambda\over\nu^2} (F_0+\half{\lambda} v^2) } -
\sqrt{ \vec k^2 + \tilde g_2{}(\vec k) {\lambda\over\nu^2}
(F_0+\half{\lambda} v_0^2) } \right\} 
\nonumber\\ 
&& + O(\A^2).
\end{eqnarray}
Remembering that $\lim_{\vec k \rightarrow \vec 0} {\tilde g_2{}}(\vec
k) = 1$, it is clear that there are no infrared divergences $(\vec k
\rightarrow \vec 0)$ to worry about, at least for this effective
potential at one loop order.  Furthermore, modulo possibly perverse
choices for the spatial noise spectrum, the tadpole term $F_0$ is
essential in renormalizing the theory. Note that the bare
potential contains terms proportional to $v^0$, $v^2$, and $v^4$,
whereas the one-loop contribution, when one expands it in powers of
$v^2$ has terms proportional to $v^{2n}$ for $n=1,2,3...$. For
definiteness, let us now take the spatial noise spectrum to be cutoff
white (the temporal spectrum has already been chosen to be exactly
white), {\em i.e.},
\begin{equation}
\tilde g_2(\vec k) =\tilde g_2(\vert \vec k \vert) 
= \Theta(\Lambda - k).
\end{equation}
The effective potential is then
\begin{eqnarray}
\label{E:effpotential}
{\cal V}[v;v_0] &=& 
\half \left[
(F_0+\half{\lambda} v^2)^2 - (F_0+\half{\lambda} v_0^2)^2  
\right] 
\nonumber\\
&&
+ \half \A \nu \int_{k < \Lambda} {\d^d \vec k |k|  \over (2\pi)^{d}} 
\left\{
\sqrt{  \vec k^2 
+  {\lambda\over\nu^2} (F_0+\half{\lambda} v^2)
} 
- \sqrt{ \vec k^2 
+  {\lambda\over\nu^2} (F_0+\half{\lambda} v_0^2)
}
\right\}
\nonumber\\
&&
+ O(\A^2).
\end{eqnarray}
With this choice of noise, the $v^2$ term is ultraviolet divergent and
proportional to $\Lambda^d$, the $v^4$ term is proportional to
$\Lambda^{d-2}$, and the $v^6$ term to $\Lambda^{d-4}$.  To have any
hope of absorbing the infinities into the bare action, thereby
permitting us to take the $\Lambda\to\infty$ limit, we must have $d<4$
(because there is no $v^6$ term in the bare potential).  That is: the
massless KPZ equation (subject to white noise) is one-loop ultraviolet
renormalizable {\em only} in 1, 2, and 3 space dimensions. (In $0$
space dimensions the KPZ equation is trivial.)  And even so, one-loop
renormalizability requires an explicit tadpole term. (Without a
tadpole term there is no term proportional to $v^2$ in the zero-loop
potential, and therefore there is no possibility of renormalizing the
leading divergence.)  Strictly speaking the claim of one-loop
renormalizability also requires investigation of the wave-function
renormalization. This appears in the one-loop effective action, which
is beyond the scope of the present paper (here we confine attention to
the one-loop effective potential), and will be discussed in future
work.

It is also clear from the above that the ultraviolet renormalizability
of the KPZ equation depends critically on the high momentum behaviour
of the noise. Let us temporarily return to equation
(\ref{E:eff-potl}), and suppose that the noise is power-law
distributed in the ultraviolet with $\tilde g_2(\vec k) = \tilde
g_2(\vert \vec k \vert) \approx (k_0/k)^{\theta} \;
\Theta(\Lambda-k)$. Then the $n$'th term in the expansion has
ultraviolet behaviour proportional to $(F_0+\half\lambda v^2)^{n}
\Lambda^{d+2-2n-n \theta}$. The massless KPZ equation is then one-loop
ultraviolet renormalizable provided the $n=3$ term (and higher terms)
are ultraviolet finite, that is, for $d<4+3\theta$.  We will not deal
with these issues any further in this paper but will instead confine
ourselves to white noise.  (Recall that it is only at intermediate
stages that the white noise is regulated by an ultraviolet spatial
cutoff. After renormalization, we will be considering noise that is
exactly white.)

We can also point out an unexpected result of this calculation: There
is a formal connection between the massless KPZ equation and massless
$\lambda (\phi^4)_{[d+1]+1}$ QFT. Consider equation
(\ref{E:effpotential}).  This is recognizable (either from QFT or
equilibrium statistical field theory) as the effective potential for
$\lambda \phi^4$ Lorentzian QFT in $d+1$ space dimensions, or
equivalently $\lambda\phi^4$ statistical field theory in $(d+1)+1$
Euclidean spacetime dimensions~\cite{Weinberg,Zinn-Justin}.  To make
the connection, interpret $\lambda F_0/\nu^2$ as the mass term $m^2$
of the QFT, $\lambda^2/\nu^2$ as $\lambda_{\mathrm QFT}$ (the coupling
constant of the QFT), and $v$ as the mean field $\phi_{\mathrm QFT}$.

Note that we have at this stage (temporarily) kept the parameters
$(F_0)_{\mathrm renormalized}$ and $v_0$ non-zero for clarity.  We now
invoke the type I and type II Galilean symmetries of the KPZ equation
to fix these parameters:

\noindent---(1) In the fluid dynamical interpretation, the tadpole
term does not have any physical significance. (Because of the symmetry
$F_0\to F_0-\kappa$, $\phi\to\phi-\kappa t$, the tadpole does not
affect any of the physical fluid velocities.)  We are interested in a
static background field configuration $\vec v_0$, but from the
equation of motion we see that for the spatially averaged field
\begin{equation}
{1\over \Omega} 
\left\langle 
\int {\partial \phi \over \partial t} \; \d^d\vec x 
\right\rangle =
F_0 + \half \lambda v_0^2 + O(\A),
\end{equation}
so we must set the {\em renormalized} value of $F_0$ to $-\half
\lambda v_0^2$.  (This works for arbitrary Gaussian noise.)  Secondly,
the type II Galilean invariance of the fluid dynamical system allows
us to set $\vec v_0 = \vec 0$ without loss of generality: pick a
coordinate system moving with the bulk fluid velocity of the
background field. (This requires translation-invariant and temporally
white noise.) Combining the two symmetries yields $(F_0)_{\mathrm
renormalized}=0$ and $v_0=0$.

\noindent---(2) In the surface growth interpretation $(F_0)_{\mathrm
renormalized}$ is a contribution to the spatially-averaged velocity of
the interface.  This average velocity term can always be scaled away
by the shift $\phi \rightarrow \phi - (F_0)_{\mathrm renormalized} \;
t$, which in the surface growth interpretation results from a type I
Galilean transformation that places one in an inertial frame moving
with the average surface profile. Doing so again sets the {\em
renormalized} value of $(F_0)_{\mathrm renormalized}$ to $-\half
\lambda v_0^2$.  Secondly, the type II Galilean invariance now
corresponds to tilting the coordinate system away from the
vertical. For any constant slope background field a simple tilt can
then be used to set the slope to zero. Combining the two symmetries
yields $(F_0)_{\mathrm renormalized}=0$ and $v_0=0$, so the surface
growth interpretation and the fluid dynamics interpretation are in
agreement as to what the physically relevant choice of parameters is.

The perhaps somewhat unusual way in which the tadpole $F_0$ is first
introduced and then renormalized to zero has a direct analog in
ordinary $\lambda \phi^4$ QFT: when we consider {\em massless}
$\lambda \phi^4$ QFT it is well known that a bare mass must be
introduced to successfully complete the renormalization program.  It
is only {\em after} renormalization that the renormalized mass may be
set to zero, and massless $\lambda \phi^4$ QFT makes sense only in
this post-renormalization fashion~\cite{Weinberg,Zinn-Justin}.

\subsection{Massless KPZ: $d=1$}

\begin{figure}[htb]
\vbox{\hfil\epsfbox{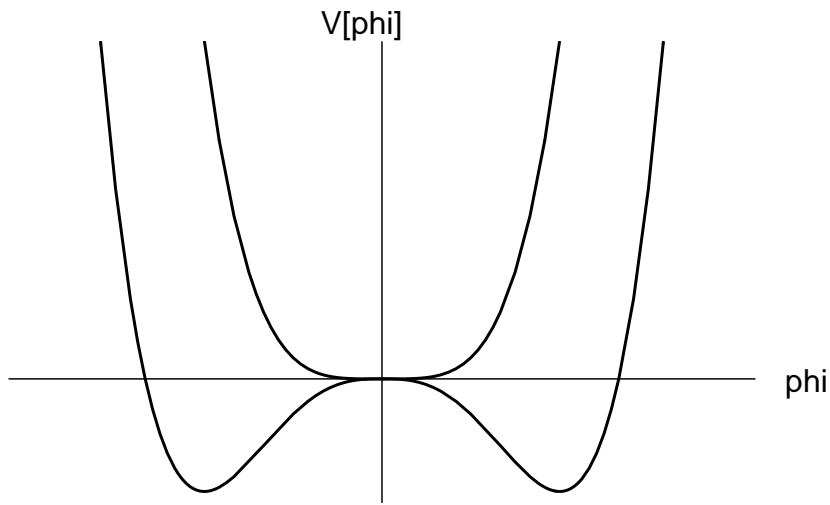}\hfil}
\caption{%
The one-loop effective potential for the KPZ equation in $d=1$ space
dimensions.  The behaviour at the origin is non-analytic in that the
third derivative is discontinuous. This distorted Mexican hat
potential indicates the onset of dynamical symmetry breaking.  For
comparison we also plot the zero-loop (``classical'') effective
potential.  Note that there is no symmetry breaking at tree level.
}\label{F:d1}
\end{figure}

For $d=1$ we are interested in the integral
\begin{equation}
\int_0^{\Lambda} \d k^2 \left[
\sqrt{k^2+ a}- \sqrt{k^2 + b}
\right].
\end{equation}
This integral is the restriction to one spatial dimension of
(\ref{E:effpotential}). It is divergent, and it will require
renormalization to extract physical answers from this formal result.
The most direct way of proceeding is via the ``differentiate and
integrate'' trick where one considers~\cite{Weinberg}
\begin{equation}
{\cal I}(a) \define 
\int_0^{\infty} \d k^2 \left[ \sqrt{k^2+ a}- \sqrt{k^2} \right].
\end{equation}
Now differentiate twice
\begin{equation}
{\d^2 \;{\cal I}(a) \over \d a^2} = 
-{1\over4}\int_0^{\infty} \d k^2 {1\over({k^2+ a })^{3/2}} = 
-{1\over 2 \sqrt{a}}.
\end{equation}
(This last integral is now well defined and finite.) If we integrate
the above equation twice, we get
\begin{equation}
{\cal I}(a) = \kappa a - {2\over3} a^{3/2}.
\end{equation}
Here $\kappa$ is a constant of integration (which happens to be
infinite). There would, in principle, be a second constant of
integration, but that is fixed to be zero by the condition that ${\cal
I}(0) =0$. We absorb $\kappa$ into the bare action, where it
renormalizes $F_0$. That is, we write
\begin{equation}
(F_0)_{\mathrm bare} = (F_0)_{\mathrm renormalized}  + 
\A \;[\delta F_0] + O(\A^2),
\end{equation}
where the counterterm $[\delta F_0]$ is chosen so as to render the
integral (when expressed in terms of renormalized quantities)
finite~\cite{Weinberg,Zinn-Justin}.
A brief calculation yields
\begin{equation}
\delta F_0 = -\frac{\kappa \lambda}{4 \pi \nu}.
\end{equation}
In one space dimension the only parameter that gets renormalized at
order $O(\A)$ is this tadpole term.  In terms of renormalized
quantities the effective potential is thus
\begin{eqnarray}
\label{E:KPZ-1}
{\cal V}[v;v_0;d=1] &=& 
\half\left[
(F_0+\half{\lambda} v^2)^2 - (F_0+\half{\lambda} v_0^2)^2 
\right]
- {1\over6\pi} \A {\lambda^{3/2} \over \nu^2} 
\left[
\left(F_0+\half{\lambda} v^2\right)^{3/2} - 
\left(F_0+\half{\lambda} v_0^2\right)^{3/2}
\right]
+ O(\A^2).
\end{eqnarray}
These are now renormalized parameters at order $O(\A)$.  Setting the
slope $v_0$ and the renormalized tadpole $F_0$ to their physical values of
zero yields
\begin{eqnarray}
{\cal V}[v;v_0=0;d=1] &=& 
{\lambda^2\over8}v^4
- {1\over6\pi} \A {\lambda^3  \over {2^{{3}/{2}} \nu^2}} 
 |v|^3
+ O(\A^2).
\end{eqnarray}
Note that the potential is not analytic at zero field (a phenomenon
well-known from massless QFTs) \cite{Weinberg,Zinn-Justin}.  For large
$v$ the classical potential dominates, while for small $v$ the
one-loop effects dominate. The minimum of the potential is not at
$v=0$, since the $-|v|^3$ term is negative and dominant near $v=0$.
Thus we encounter something very interesting ---the system undergoes
{\em dynamical symmetry breaking} (DSB) in a manner qualitatively
similar to the Coleman--Weinberg mechanism of particle physics. The
qualitative form of the effective potential is sketched in
figure~\ref{F:d1}.

Symmetry breaking is said to be spontaneous if there is a symmetry in
the potential that is not shared by the zero-loop ground states ({\em
e.g.}, the Higgs mechanism). If the symmetry is preserved at the
classical level (zero-noise), but is broken once fluctuations are
taken into account (broken by loop effects) then the symmetry breaking
is said to be dynamical ({\em e.g.}, massless $\lambda \phi^4$ theory,
Coleman--Weinberg mechanism).

{For} small $\vert \vert \, \vec v \, \vert \vert$ the one-loop
contributions drive the minimum of the potential away from zero field.
To find the location of the minimum we calculate
\begin{equation}
{\d{\cal V}[v;v_0=0;d=1]\over \d v} =
\half \lambda^2 v^3 
-{1\over2\pi}\;\A\;{\lambda^3 \over {2^{{3}/{2}} \nu^2}}  
\sign(v)\, v^2 
+ O(\A^2) = 0.
\end{equation}
This permits us to {\em estimate} the shift in the expectation value
of the velocity field
\begin{equation}
v_{\mathrm min} = 
\pm \A \; {\lambda\over2\pi \; 2^{1/2} \; \nu^2} + O(\A^2).
\end{equation}
Unfortunately, the presence of the unknown $O(\A^2)$ terms renders it
impossible to make any definitive statement about the precise value of
$v_{\mathrm min}$, apart from the fact that it is
non-zero~\cite{Rivers,Pokorsky}. (This is a common feature in DSB, as
perturbatively detecting the occurrence of DSB is easier than finding
the precise location of the minimum.) To complete the specification of
$v_{\mathrm min}$ one would have to calculate the $O(\A^2)$ terms in
the effective potential and verify that the one-loop estimate of
$v_{\mathrm min}$ occurs at values of the velocity where the $O(\A^2)$
term is negligible. This is not a trivial task, and we refer the
reader to several texts where this is more fully
addressed~\cite{Rivers,Pokorsky}.

This DSB is particularly intriguing in that it suggests the
possibility of a noise driven pump. For example, in thin pipes where
the flow is essentially one-dimensional, and provided the physical
situation justifies the use of the vorticity-free Burgers equation for
the fluid, this result indicates the presence of a bimodal instability
leading to the onset of a bulk fluid flow with velocity dependent on
the noise amplitude. In the surface growth (line growth)
interpretation the onset of DSB corresponds to an initially flat line
breaking up into a sawtooth pattern of domains in which the slope
takes on the values $\pm v_{\mathrm min}$.

\subsection{Massless KPZ: $d=2$}

The present discussion is relevant to either (1) surface evolution on
a two dimensional substrate, or (2) thin superfluid films (since
superfluids are automatically vorticity-free, justifying the
application of the zero-vorticity Burgers equation).

An immediate consequence of the analogy between the one-loop effective
potential for the massless KPZ equation (for white noise) and that for
the massless $\lambda \phi^4$ QFT is that we can write down the
renormalized one-loop effective potential for $d=2$ (space dimensions)
by inspection, merely by recalling that for the $d=4$ (spacetime
dimensions) scalar field theory we have (see, {\em e.g.,}
\cite{Weinberg,Zinn-Justin})
\begin{eqnarray}
{\cal V}[v;v_0;d=2] &=& 
\half\left\{ 
[F_0(\mu)+\half{\lambda(\mu)} v^2]^2 - 
[F_0(\mu)+\half{\lambda(\mu)} v_0^2]^2 
\right\} 
\nonumber\\
&&
+\half \A  {1  \over (2\pi)^{2}} 
{\lambda^2\over\nu^3}
\Bigg\{
[F_0(\mu)+\half{\lambda(\mu)} v^2]^2 
\ln\left[{F_0(\mu)+\half{\lambda(\mu)} v^2\over \mu^2 }\right]
\nonumber\\
&&
\qquad\qquad\qquad
-[F_0(\mu)+\half{\lambda(\mu)} v_0^2]^2 
\ln\left[{F_0(\mu)+\half{\lambda(\mu)} v_0^2\over \mu^2 }\right]
\Bigg\}
+ O(\A^2).
\end{eqnarray}
Here $\mu$ is the renormalization scale, as normally used in QFT
\cite{Weinberg,Zinn-Justin}.  (In a condensed matter setting this
might be thought of as a measure of the coarse-graining
scale.)  Its presence is a side effect of the logarithmic divergences,
which happen to occur in $d=2$ space dimensions for the KPZ equation.

\begin{figure}[htb]
\vbox{\hfil\epsfbox{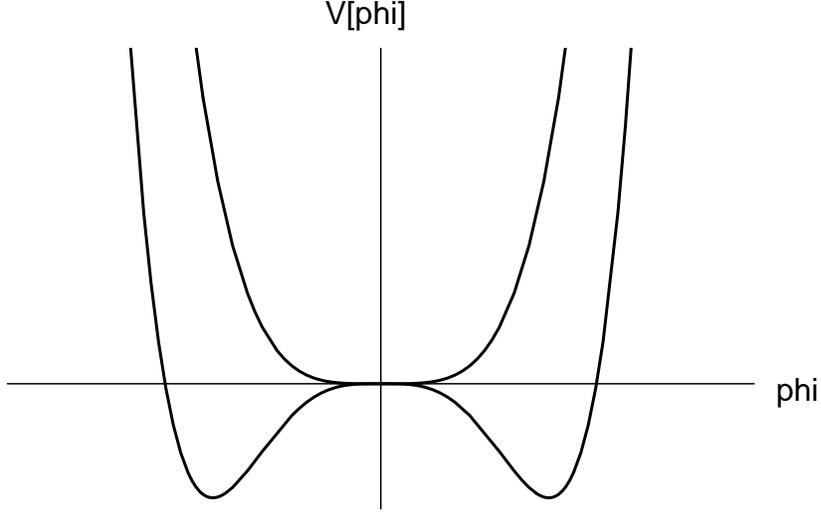}\hfil}
\caption{%
The one-loop effective potential for the KPZ equation in $d=2$ space
dimensions.  The behaviour at the origin is non-analytic in that the
fourth derivative exhibits a logarithmic singularity. This distorted
Mexican hat potential indicates the onset of dynamical symmetry
breaking. For comparison we also plot the zero-loop (``classical'')
effective potential, note that there is no symmetry breaking at tree
level.  }\label{F:d2}
\end{figure}

If we now tune $v_0$ and the renormalized value of the tadpole $F_0$
to their physical values of zero we obtain
\begin{eqnarray}
{\cal V}[v;v_0=0;d=2] &=& 
{\lambda^2\over 8} v^4 
+\half \A  {1  \over (2\pi)^{2}} 
{\lambda^4\over4\nu^3}
v^4 
\ln\left({v^2\over \mu^2 }\right)
+ O(\A^2).
\end{eqnarray}
This potential is zero at $v=0$, then becomes negative, though for
large enough fields ($v>\mu$) the potential again becomes positive.
That this potential has a nontrivial minimum exhibiting DSB is exactly
the analog in this massless KPZ problem of the well-known
Coleman--Weinberg mechanism encountered in the extensions of the
standard model of particle physics~\cite{Weinberg,Zinn-Justin}.  As in
$d=1$, it is easy to see that the minimum effective potential occurs
for $v_{\mathrm min} \neq 0$, but because of the presence of unknown
$O(\A^2)$ terms it is difficult to give a good estimate for the value
of $v_{\mathrm min}$~\cite{Rivers,Pokorsky}.  The qualitative form of
the effective potential is sketched in figure~\ref{F:d2}. If we
estimate the location of the minimum by differentiating the effective
potential we obtain
\begin{equation}
v_{\mathrm min} = \pm \mu \;
\exp\left( 
-{(2\pi)^2\;\nu^3\over2\;\lambda^2 \;\A} - {1\over4}  +O(\A) 
\right).
\end{equation}
Note that $v_{min} \to 0$ as $\A \to 0$, as it should, to recover the
tree level minimum $v_{min}=0$.

Following the analysis of~\cite{Gato} we can immediately extract the
one-loop beta function. In order to do so, we make use of the fact
that the bare effective potential does not depend on the
renormalization scale:
\begin{equation}
{\mu \; \d \over \d\mu} \; {\cal V}[v;v_0;d] =0.
\end{equation}
We get
\begin{equation}
\label{E:beta1}
{\mu \; \d \over \d\mu} \; 
\left( 
{\lambda^2(\mu)\over 8} v^4
\right)
=
{\A\over16\pi^2} {\lambda^4 v^4\over\nu^3} + O(\A)^2.
\end{equation}
Comparing the coefficients of the $v^4$ terms we obtain
\begin{equation}
\label{E:beta2}
\beta_\lambda \define {\mu \; \d \over \d\mu} \; \lambda = 
{\A \over4\pi^2} {\lambda^3\over\nu^3} + O(\A)^2.
\end{equation}
We cannot extract the beta function for the wavefunction
renormalization of $\phi$ (or $v$) from the present analysis. This
would require a calculation of the effective action for an
inhomogeneous field, an issue which we postpone for the
future. Equations (\ref{E:beta1}) and (\ref{E:beta2}) are correct
because one can show~\cite{HMPV-kernel} that to one loop there is no
wavefunction renormalization for the KPZ field in this background.

\subsection{Massless KPZ: $d=3$}

\begin{figure}[htb]
\vbox{\hfil\epsfbox{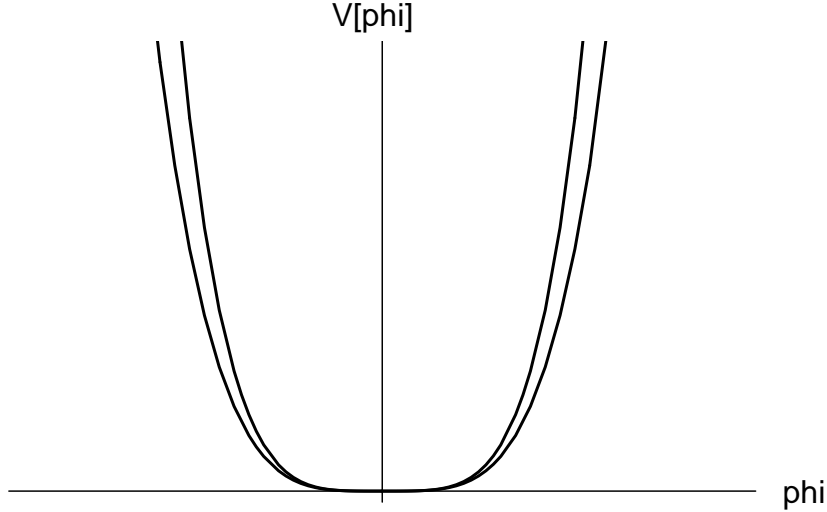}\hfil}
\caption{%
The one-loop effective potential for the KPZ equation in $d=3$ space
dimensions.  The behaviour at the origin is non-analytic in that the
fifth derivative is discontinuous. The fact that the potential has
only one minimum indicates that there is no symmetry breaking.  The
zero-loop (``classical'') potential is also plotted for comparison
purposes. In $d=3$ space dimensions one-loop effects are not dramatic,
and the qualitative form of the effective potential is not
significantly altered by one loop effects.  }\label{F:d3}
\end{figure}

This case is of physical interest for three-dimensional fluids.  For
$d=3$ we are interested in the integral
\begin{equation}
\int_0^\infty \d k^2 k^2 \left[
\sqrt{k^2+ a }- \sqrt{k^2 + b}
\right] .
\end{equation}
Define the quantity
\begin{equation}
{\cal I}(a) \define 
\int_0^{\infty} \d k^2 k^2 \left[ \sqrt{k^2+ a}- \sqrt{k^2} \right].
\end{equation}
The ``differentiate and integrate'' trick~\cite{Weinberg} leads to
\begin{equation}
{\d^3 \;{\cal I}(a) \over \d a^3} = 
{3\over8}\int_0^{\infty} \d k^2 {k^2\over({k^2+ a})^{5/2}} = 
{1\over 2 \sqrt{a}}.
\end{equation}
(The integral is now well behaved and finite). We now integrate this
thrice to obtain
\begin{equation}
{\cal I}(a) = \kappa_1 a + \kappa_2 a^2 + {4\over15} a^{5/2}.
\end{equation}
Here $\kappa_1$ and $\kappa_2$ are now two constants of integration
(which happen to be infinite). There would in principle be a third
constant of integration, but that is fixed to be zero by the condition
that ${\cal I}(0) =0$. (The sign in front of the $a^{5/2}$ term is
important, since this sign is positive we will see that there is no
possibility of DSB in three space dimensions.) We absorb $\kappa_1$
and $\kappa_2$ into the bare potential, where they renormalize both
$F_0$ and $\lambda$. The relevant counterterms are
\begin{equation}
\delta F_0 = 
-{\kappa_1\over8\pi^2} {\lambda\over\nu}\; , \; \; \; \; \; 
{\rm and} \; \; \; \; \; 
\delta \lambda =
- {\kappa_2\over4\pi^2} \left({\lambda\over\nu}\right)^3
\; ,
\end{equation}
so as to yield
\begin{eqnarray}
{\cal V}[v;v_0;d=3] &=& 
\half\left[
(F_0+\half{\lambda} v^2)^2 -  
(F_0+\half{\lambda} v_0^2)^2 
\right]
+ {1\over 30\pi^2}\A {\lambda^{5/2} \over \nu^4}
\left[
\left( F_0+\half{\lambda} v^2\right)^{5/2} - 
\left( F_0+\half{\lambda} v_0^2 \right)^{5/2}
\right]
\nonumber\\
&&
+ O(\A^2).
\end{eqnarray}
These are now all renormalized parameters at order $O(\A)$.  Setting
to zero both $v_0$ and the renormalized value of $F_0$ we have
\begin{eqnarray}
{\cal V}[v;v_0=0;d=3] &=& 
{\lambda^2\over8} v^4
+ {1\over 30\pi^2}\A {\lambda^{5} \over 2^{{5}/{2}} \nu^4}  |v|^5
+ O(\A^2).
\end{eqnarray}
At zero-loops the vacuum is symmetric at $v=0$. Adding one-loop
physics does not change this. Note that there is an all-important sign
change in the one-loop contribution to the effective potential in
comparing $d=3$ with $d=1$. The DSB that is so interesting in $d=1$,
and via the Coleman--Weinberg mechanism in $d=2$, is now completely
absent in $d=3$. Observe also that the effective potential is
non-analytic at $v=0$. The qualitative form of the effective potential
is sketched in figure~\ref{F:d3}.

\section{Discussion}

In this paper we have explicitly calculated the renormalized one-loop
effective potential for the massless KPZ equation in 1, 2, and 3 space
dimensions. Although the effective potential is by definition
time-independent, we already find a very interesting structure for the
static ground states of the system. There is a close analogy between
the statics of the massless KPZ system and the static behaviour of
massless $\lambda \phi^4$ QFT---and much of the vacuum structure of
the massless $\lambda \phi^4$ QFT carries over into the ground state
structure of the massless KPZ equation. It is important to underscore
the fact that the analysis presented here has focussed on the short
distance or ultraviolet properties of the KPZ equation, which as we
have seen, reveal a complementary phenomenology to the perhaps more
common studies concerned with the infrared, or long distance
properties of the KPZ equation. This distinction shows up, among other
places, in the requirement of the bare tadpole term, the structure of
the ultraviolet divergences requiring renormalization, the phenomena
of dynamical symmetry breaking, and the marked dimension dependence of
the one-loop ultraviolet renormalization group equation, and its
associated fixed point.  The generality of these structural and
dimension-dependent features are confirmed by studying the (one-loop)
effective action associated to the KPZ equation; these general results
will be presented elsewhere.

In 1 and 2 space dimensions we have exhibited the occurrence of
dynamical symmetry breaking.  In the hydrodynamic interpretation
symmetry breaking corresponds to instability of the zero-velocity
background, leading to the onset of bulk flows in the fluid. In the
surface growth interpretation symmetry breaking corresponds to
instability of the planar interface, leading to a domain structure
wherein different domains exhibit different slopes (all of the same
magnitude).  Thus even the static ground state structure of the KPZ
equation is surprisingly rich, considerably richer than one might have
reasonably expected. These considerations lead one to conjecture that
the one-loop methods developed in \cite{HMPV-spde} and illustrated
here, may be profitably applied to reveal the onset of instabilities
in the Kuramoto--Sivashinki (KS) equation. The KS equation, used to
model flame-front propagation, may be regarded as a generalization of
the KPZ equation with an additional fourth-derivative term $(\vec
\nabla^4 \phi)$. In two dimensional combustion, recent experiments
have demonstrated the existence of so-called ``fingering''
instabilities as a function of oxygen flow across the surface
\cite{Fingering}. On the other hand, there is no fingering in
three-dimensional combustion (because of convection).  The KS equation
is also Galilean invariant. Thus, it may well exhibit a dimension
dependent one-loop DSB akin to that of the KPZ equation.

\section*{Acknowledgments}

In Spain, this work was supported by the Spanish Ministry of Education
and Culture and the Spanish Ministry of Defense (DH and JPM). In the
USA, support was provided by the US Department of Energy (CMP and MV).
The research of CMP is supported in part by the Department of Energy
under contract W-7405-ENG-36.  Additionally, MV wishes to acknowledge
support from the Spanish Ministry of Education and Culture through the
Sabbatical Program, to recognize the kind hospitality provided by
LAEFF (Laboratorio de \Astrofisica\ Espacial y \Fisica\ Fundamental;
Madrid, Spain), and to thank Victoria University (Te Whare Wananga o
te Upoko o te Ika a Maui; Wellington, New Zealand) for hospitality
during final stages of this work.

\appendix
\section{Jacobian functional determinant for Burgers/KPZ}

We are interested in evaluating the Jacobian determinant for the {\em
massive} KPZ equation, with $F[\phi(\vec x)] = F_0 -\nu m^2 \phi +
{\lambda\over2} (\vec \nabla\phi)^2$:
\begin{equation}
{\cal J} = \det\left( D - {\delta F\over\delta\phi} \right).
\end{equation}
We do so by means of the formalism developed in~\cite{HMPV-spde}. Thus
\begin{equation}
{\delta F[\phi(\vec x)]\over\delta\phi(\vec y)} = 
-\nu \; m^2 \; \delta(\vec x-\vec y) 
+ \lambda \; 
\vec\nabla \phi(\vec x) \cdot \vec \nabla_{\vec x} \; 
\delta(\vec x-\vec y).
\end{equation}
And so
\begin{eqnarray}
\Tr\left[
 {\delta F[\phi(\vec x)]\over\delta\phi(\vec y)}
\right] 
&=&
\Tr\left[
-\nu \; m^2 \; \delta(\vec x-\vec y) 
+\lambda \;  \vec\nabla \phi \cdot \vec \nabla \delta(\vec x-\vec y) 
\right]
\\
&=&
-\nu \; m^2 \; \delta^d(\vec 0) 
- \lambda \; \delta^d(\vec 0) \int \vec \nabla^2 \phi \; \d x
\\
&=&
- \delta^d(\vec 0) \left\{ \nu m^2 + \lambda\; 
\int_{\cal B} \vec \nabla \phi \cdot d \vec S \; \d t \right\}
\\
&=&\hbox{field independent constant}
\\
&=& 0?
\end{eqnarray}
The ante-penultimate expression is somewhat formal, but is certainly
field independent, with at worst some dependence on the boundary
conditions.  Invoking the formalism developed in~\cite{HMPV-spde},
this implies that ${\cal J}_{\mathrm (KPZ)}$ is a field-independent
constant. That is, for either the massless or massive KPZ equation,
the functional Jacobian determinant is a field-independent constant
which can be ignored.  There are general formal arguments (see for
example Zinn-Justin~\cite{Zinn-Justin} pp. 373, 307, or related
comments in Itzykson--Zuber~\cite{Itzykson-Zuber} p. 448) to the
effect that terms proportional to $\delta^d(\vec 0)$ can always be
discarded in dimensional regularization. 

\section{Feynman rules for Burgers/KPZ field theory}

{From} the massive Burgers/KPZ stochastic differential equation
\begin{equation}
\left[{\partial\over\partial t} - \nu (\vec\nabla^2-m^2)\right] \phi = 
F_0 + {\lambda\over2} (\vec \nabla\phi)^2 + \eta,
\end{equation}
we deduce the partition function~\cite{HMPV-spde}:
\begin{eqnarray}
Z[J] &=& 
\int ({\cal D} \phi)\;
\exp\left( \int J \phi \right)
\nonumber\\
&&\qquad
\exp\left( -{1\over2} \int \int
\left[
\partial_t \phi - \nu(\vec \nabla^2-m^2) \phi 
-F_0 - {\lambda\over2} (\vec \nabla\phi)^2
\right] 
 G_\eta^{-1}
\left[
\partial_t \phi - \nu(\vec \nabla^2-m^2) \phi 
-F_0 - {\lambda\over2} (\vec \nabla\phi)^2
\right]
\right).
\end{eqnarray}
We point out at this stage, that the functional determinant can be
discarded in this case.  There is only one propagator and two
vertices:

\begin{eqnarray}
&&\hbox{Propagator}:
\nonumber
\\
&&
\qquad
G_{\mathrm field}(\vec k,\omega)= 
{ G_\eta(\vec k,\omega)\over \omega^2 + \nu^2 (\vec k^2+m^2)^2};
\\
&&
\hbox{Three-point vertex:} 
\nonumber
\\
&&
\qquad
[F_0 - \half\lambda \;
(\vec k_1 \cdot \vec k_2 )] {1\over (2\pi)^{d+1}} \;
{(-i\omega_3+\nu \vec k_3^2)\over G_\eta(\vec k_3,\omega_3) } \; 
\delta(\vec k_1 + \vec k_2 + \vec k_3) \; 
\delta(\omega_1 + \omega_2 + \omega_3); 
\\
&&
\hbox{Four-point vertex:}
\nonumber
\\
&&  
\qquad
{1\over2}\; 
{[F_0-\half\lambda(\vec k_1 \cdot \vec k_2 )]
 [F_0-\half\lambda(\vec k_3 \cdot \vec k_4 )]
\over(2\pi)^{d+1}\;  G_\eta(\vec k_1+\vec k_2,\omega_1+\omega_2)} \; 
\delta(\vec k_1 + \vec k_2 + \vec k_3 + \vec k_4) \;
\delta(\omega_1 + \omega_2 + \omega_3 + \omega_4).
\end{eqnarray}
Feynman diagram computations (in this ``direct'' formalism) are now
straightforward (though tedious!). If one is careful to ask only
physical questions, calculations based on this ``direct'' formalism
will yield the same answers as those extracted from Feynman diagrams
based on the more common Martin--Siggia--Rose (MSR)
formalism~\cite{Zinn-Justin,MSR,De-Dominicis-Peliti}.


\end{document}